\documentclass{article}

\usepackage[preprint]{neurips_2026}

\usepackage[utf8]{inputenc} 
\usepackage[T1]{fontenc}    
\usepackage{hyperref}       
\usepackage{url}            
\usepackage{booktabs}       
\usepackage{amsfonts}       
\usepackage{nicefrac}       
\usepackage{microtype}      
\usepackage{xcolor}         
\usepackage{mathtools}
\usepackage{amsmath}
\usepackage{amsthm}
\usepackage{physics}
\usepackage{quantikz}
\usepackage{algorithm}
\usepackage{algpseudocode}

\newtheorem{proposition}{Proposition}
\newtheorem{corollary}{Corollary}
\newtheorem{theorem}{Theorem}
\newtheorem{definition}{Definition}

\newtheorem{lemma}{Lemma}

\title{A New Framework for Quantum Reinforcement Learning}

\author{
  Alireza Tavakoli\textsuperscript{1}\thanks{Corresponding author.}\qquad
  Mohsen Heidari\textsuperscript{1}\\\\
  \textsuperscript{1}Indiana University\\
  \texttt{\{altavak,mheidar\}@iu.edu}
}

\begin{document}

\maketitle

\begin{abstract}

    Quantum reinforcement learning (QRL) has emerged as a promising approach for controlling quantum systems and accelerating learning tasks with quantum resources. However, most existing QRL formulations still retain a classical reinforcement learning interface: states and actions are often encoded in computational bases and rewards are extracted through measurements. This interface is restrictive for genuinely quantum environments: choosing a measurement can discard entanglement information, while repeated intermediate measurements collapse the evolving state and require costly re-preparation. 
    This motivates a fully QRL framework in which the agent, environment, rewards, and return accumulation are all quantum. 
    In this paper, we take a step toward this goal and introduce a new QRL framework in which both the policy and the environment are modeled by \emph{conditional quantum channels}. 
    We then propose a quantum policy-gradient algorithm for parameterized conditional quantum channels with a trainable control basis. 
    We also derive finite- and infinite-horizon gradient estimators using a new parameter-shift rule.
    This rule is needed because policy parameters are shared across repeated interaction blocks and the infinite-horizon objective is not a finite circuit; rather than shifting every occurrence and evaluating the whole trajectory, our estimator shifts only one sampled policy block.
    Sampling this block from a geometric distribution yields an unbiased gradient estimator under bounded rewards.
    Finally, we establish convergence of the resulting stochastic gradient-ascent procedure to stationary points under standard smoothness and step-size assumptions.
\end{abstract}

\section{Introduction}
Reinforcement learning (RL) provides a general framework for sequential decision-making, in which an agent learns to choose actions by interacting with an environment to maximize long-term rewards. It has become a standard tool for control and optimization in complex domains where explicit modeling is difficult, ranging from robotics and games \citep{Kober2013IJRR, mnih2013playingatarideepreinforcement, pinto2017asymmetricactorcriticimagebased, tang2024deepreinforcementlearningrobotics} to recommendation systems and healthcare \citep{Li_2010, jayaraman2024primer}.

Quantum systems are a particularly compelling setting for RL. Their dynamics are high-dimensional, noisy, partially observed, and often difficult to model accurately, while many operational tasks, such as qubit calibration, require adaptive feedback under hardware drift and limited measurement access. These features make model-free, interaction-driven methods especially attractive: rather than relying on a complete Hamiltonian or noise model, an RL agent can learn control policies directly from experimental feedback. This perspective has already led to successful applications across quantum technologies, including model-free qubit calibration and gate design, error-robust quantum gate tuning, quantum-enhanced learning protocols, and RL control of quantum error correction \citep{shindi2023modelfreequantumgatedesign, Baum_2021, Saggio_2021, sivak2025reinforcementlearningcontrolquantum}.

However, most of these approaches are hybrid classical-quantum models in which a classical agent repeatedly queries a quantum device, extracts classical rewards through measurements, and updates classical parameters. This interface is costly and restrictive as measurements cause state collapses and hence require repeated state preparation. 
In addition, variational quantum policies trained through classical optimization can suffer from barren plateaus \citep{McClean2018, Ragone2024}; and in domains such as quantum chemistry \citep{WEIDMAN2024102105} and many-body physics \citep{Fauseweh2024}, the environment itself may be a quantum system whose full state reconstruction requires sample-intensive tomography \citep{Huang_2020, King_2025}. 

These limitations motivate a fully quantum reinforcement learning (QRL) formulation in which the agent, the reward, and the environment are all quantum.
Rather than measuring the device after each step and passing classical outcomes to a classical optimizer, a fully quantum agent can apply a policy channel, interact with a quantum environment, accumulate reward information in auxiliary registers, and defer measurements until the evaluation of the overall objective. The framework is relevant for applications in coherent feedback control \citep{PhysRevA.62.022108}, quantum chemistry \citep{Peruzzo2014}, and many-body quantum simulation \citep{Georgescu_2014}, where the agent may need to optimize behavior over quantum trajectories that are classically inaccessible.

Recent results in quantum bandit algorithms provide evidence that quantum access can improve RL primitives, but they do not model a full coherent state-action-reward trajectory \citep{Wang_2021,wan2022quantummultiarmedbanditsstochastic}.
Several other existing works formulate QRL frameworks. 
For instance, \citep{dunjko2015frameworklearningagentsquantum, Dunjko_2016, Dunjko_2017} consider oracular QRL frameworks. These frameworks are suitable for speedups in classical RL, but it may not be applicable in fully quantum environments, where the relevant information may live in an unknown basis.

Coherent control of unitary and noisy quantum channels has been studied
previously~\citep{Vanrietvelde_2021,Abbott2020communication}.
\citep{PhysRevA.90.032311} study planning in quantum
partially observable Markov decision processes, where an agent selects
quantum operations. Our model instead specifies separate state, action,
instantaneous-reward, and stored-reward registers, and studies learning
a state-conditioned policy while deferring reward readout. Our
contributions are this QRL interaction model, its trainable-basis policy,
and the sampled-block gradient estimator; coherent control itself is
established prior work.

\paragraph{Our contributions.} In this work, we take a step further and develop a new fully QRL framework in which the agent and environment processes are embedded in a single model. The interaction is organized through separate state, action, reward, and reward-accumulation registers, preserving the causal structure of RL while enforcing a quantum separation of roles: the agent selects actions without directly modifying the state, and the environment updates the state and reward without directly modifying the action. Since rewards and states are not measured at intermediate steps, the trajectory can remain entangled until the final objective evaluation, avoiding repeated measurement-induced collapse and state re-preparation. The framework also supports general finite- and infinite-horizon interactions beyond purely oracle settings.

One of the main obstacles to fully QRL is that the agent is not permitted to measure the environment state or the reward signal directly, since such measurements would collapse the quantum state, disturb the quantum evolution, and destroy the very coherence the model aims to exploit. Moreover, the agent should influence the environment only through an action register, not by acting directly on the state itself; otherwise, the problem ceases to be RL and becomes direct quantum state control. In our framework, this difficulty is addressed by defining conditional quantum channels (CQCs). 
These channels modify a target register by conditioning on a control register. 
This separation preserves the agent-environment distinction and prevents either side from arbitrarily modifying the other's internal register. 
CQCs generalize controlled unitaries by allowing the operation applied to the target register to be a general quantum channel, rather than only a reversible unitary transformation.

As a motivation, quantum approximate optimization algorithms (QAOAs) can be cast as a QRL in our framework. The agent can be viewed as selecting the angles or durations of alternating Hamiltonian evolutions, while the evolving quantum state of the device plays the role of the environment; indeed, prior work has already shown that choosing QAOA parameters can be cast as an RL problem and that RL-based methods can assist QAOA optimization \citep{Patel_2024, yao2020policygradientbasedquantum, Yao_2021, Wauters_2020}. From this perspective, our framework provides a principled way to model such adaptive control directly within a fully quantum interaction loop.

We instantiate the policy class using conditional parameterized quantum circuits (CPQCs). A CPQC is a parameterized CQC in which the conditional basis is trainable. Thus, the policy learns not only the action-generating transformation, but also the basis in which the state should control the action. 
We then derive policy-gradient estimators for this CPQC policy in both finite- and infinite-horizon settings using a new parameter-shift rule.

The standard parameter-shift rule applies to fixed finite circuits and differentiates a specified gate occurrence. Our setting is different because the same policy parameters are shared across all repeated interaction blocks, so the gradient contains contributions from many time locations. In the infinite-horizon case, this becomes an infinite discounted sum and the interaction cannot be represented as a single finite circuit. Our parameter-shift rule resolves this by sampling the time location of the shifted policy block, shifting only that block, and estimating the corresponding suffix return. This avoids shifting every occurrence of the shared parameter while still producing an unbiased gradient estimator.

Finally, we analyze the convergence of the resulting procedure. Under standard smoothness, step-size, and estimation-accuracy conditions, our algorithm yields almost-sure summability of the squared true-gradient norm along the optimization trajectory. When the estimator noise is asymptotically exact, as can be achieved by increasing the estimation accuracy during training,
the stochastic gradient estimator itself converges to zero almost surely. Thus, despite the non-convex
CPQC landscape, the method converges to stationary behavior.

\section{Preliminaries}

\subsection{Reinforcement learning}\label{subsec:rl}

We review the basic elements of RL. Let \(S\) be the state space of the environment and \(A\) the action space of the agent. The agent's behavior is specified by a policy, namely a conditional distribution \(\pi_\theta(a\mid s)\) over actions given the current state, parameterized by \(\theta\).

At each time step \(t\), the agent observes the current state \(s_t \in S\), selects an action \(a_t \in A\) according to \(\pi_\theta(\cdot \mid s_t)\), receives a reward \(r_t \in \mathbb{R}\), and the environment evolves to a new state \(s_{t+1}\). In the fully observable case, the environment is modeled as a Markov decision process (MDP) \((S, A, P, R, \gamma)\), where \(P(s'\mid s, a)\) specifies the state transition law, \(R(r\mid s, a, s')\) specifies the reward distribution, and \(\gamma \in [0, 1)\) is the discount factor. The standard objective is to maximize the expected discounted return
\[
    J(\theta)=\mathbb{E}_{\pi_\theta}\!\left[\sum_{t=0}^{T-1}\gamma^t r_t\right]
\]
in the finite-horizon case. In the infinite-horizon case, \(T = \infty\) and the summation runs to infinity.

\subsection{Quantum states, observables, and channels}\label{subsec:quantum-state-observables-channels}

Let \(\mathcal{H}\) be a finite-dimensional Hilbert space associated with a quantum system. 
A quantum state on \(\mathcal{H}\) is represented by a density operator \(\rho\), that is, a positive semidefinite linear operator acting on \(\mathcal{H}\) with \(\Tr(\rho)=1\). 
Pure states are rank-one projectors of the form \(\rho = \dyad{\psi}{\psi}\), while mixed states are convex combinations of pure states.

An observable \(O\) is a Hermitian operator acting on \(\mathcal{H}\). Given a state \(\rho\), the expectation value of \(O\) is
\(
    \expval{O}_\rho \coloneqq \Tr(O\rho).
\)

The most general evolution of a quantum system is described by a quantum channel, namely a completely positive trace-preserving (CPTP) linear map \(\mathcal N\) from operators on the input Hilbert space \(\mathcal{H}_\mathrm{in}\) to operators on the output Hilbert space \(\mathcal{H}_\mathrm{out}\).
Every such channel admits a Kraus representation
\(
    \mathcal{N}\left(\rho\right) = \sum_x F_x\rho F_x^\dagger,
\)
where the Kraus operators satisfy the completeness relation
\(
    \sum_x F_x^\dagger F_x = I.
\)
For more details, see \citep{wilde2017quantum}.

\section{Fully quantum reinforcement learning framework}
\label{sec:framework}

We formulate the QRL framework as follows.
Let \(S\) be the state register, \(A\) the action register, \(R\) the instantaneous reward register, and \(C\) an auxiliary register used to store reward information over time. The auxiliary register's structure is not fixed a priori and may depend on the choice of reward-accumulation map. The interaction starts from an initial state \(\rho^{(0)}\). In our model, the environment and the agent are formulated using a specific type of quantum channel called a CQC.

In RL, the agent chooses an action based on the current state, and the environment updates the state and reward based on that action. CQCs represent these conditional updates without measuring the state or action registers. An arbitrary joint channel could allow the agent to modify the environment state directly, whereas measurement-based conditioning would collapse superpositions. Unlike controlled unitaries, CQCs also permit general CPTP updates needed to model noisy or irreversible environment dynamics.

\begin{definition}[Conditional quantum channel]
        A CQC \(\mathcal N\) acting on a control register \(X\) and a target register \(Y\) is specified by a projective resolution \(\{P_i\}_i\) on \(X\), satisfying
        \[
            P_i=P_i^\dagger, \qquad
            P_iP_j=\delta_{ij}P_i, \qquad
            \sum_i P_i=I_X,
        \]
        together with, for each \(i\), a Kraus family \(\{F_{i,x}\}_x\) on \(Y\) satisfying
        \[
            \sum_x F_{i,x}^{\dagger}F_{i,x}=I_Y .
        \]
        The same index set \(x\) is used for all \(i\), padding Kraus families with zero operators if necessary. The CQC is the channel defined as
        \[
            \mathcal N(\rho_{XY})
            =
            \sum_x K_x\rho_{XY}K_x^\dagger,
            \qquad
            K_x \coloneqq \sum_i P_i\otimes F_{i,x}.
        \]

\end{definition}
        When the control register is in the subspace selected by \(P_i\), the target register evolves according to
        \(\mathcal N_i(\sigma)=\sum_x F_{i,x}\sigma F_{i,x}^\dagger\). The control register determines which target channel is applied, but it is not measured or overwritten.

\begin{proposition}
    A CQC is a valid quantum channel.
\end{proposition}
\begin{proof}
    The map is completely positive because of its Kraus operators \(K_x\). It remains to verify trace preservation. 
    \[
        \sum_x K_x^\dagger K_x
        =
        \sum_x
        \sum_{i,j}
        P_iP_j\otimes F_{i,x}^\dagger F_{j,x}
        =
        \sum_i P_i\otimes
        \sum_x F_{i,x}^\dagger F_{i,x}
        =
        \sum_i P_i\otimes I_Y
        =
        I_{XY}.
    \]
\end{proof}

Many quantum operations, such as oracles and controlled unitaries, are special cases of CQCs.
\begin{proposition}
    Oracles are special cases of CQCs.
\end{proposition}
\begin{proof}
    A standard oracle has the form
    \(
        U_f=\sum_i \dyad{i}{i}\otimes U_{f(i)},
    \)
    where each \(U_{f(i)}\) is unitary. Choosing \(P_i=\dyad{i}{i}\) and the single Kraus operator \(F_{i,1}=U_{f(i)}\) in the CQC definition gives \(K_1=U_f\), and hence the oracle channel \(\rho\mapsto U_f\rho U_f^\dagger\).
\end{proof}

\subsection{Policy, environment, and reward accumulation channels}
\label{subsec:policy-env-reward}

\begin{figure}
    \centering
    \begin{quantikz}[row sep=0.25cm, column sep=0.1cm]
    \lstick{\(A\)} 
        & \gate{\Pi_{\theta}^{(0)}} & \qw
        & \ctrl{1} & \qw
        & \qw & \qw
        & \push{\hspace{0.05cm}\cdots\hspace{0.05cm}} & \qw
        & \gate{\Pi_{\theta}^{(T-1)}} & \qw
        & \ctrl{1} & \qw 
        & \qw & \qw \\
    \lstick{\(S\) } 
        & \ctrl{-1} & \qw
        & \gate[wires=2]{\mathcal{E}^{(0)}} & \qw
        & \qw & \qw
        & \push{\hspace{0.05cm}\cdots\hspace{0.05cm}} & \qw
        & \ctrl{-1} & \qw
        & \gate[wires=2]{\mathcal{E}^{(T-1)}} & \qw
        & \qw & \qw \\
    \lstick{\(R\)} 
        & \qw & \qw
        & \ghost{\mathcal{E}^{(0)}} 
        & \gate[wires=2]{\mathcal{G}^{(0)}} & \qw
        & \qw
        & \push{\hspace{0.05cm}\cdots\hspace{0.05cm}} & \qw
        & \qw & \qw
        & \ghost{\mathcal{E}^{(T-1)}} 
        & \gate[wires=2]{\mathcal{G}^{(T-1)}} & \qw
        & \qw \\
    \lstick{\(C\)} 
        & \qw & \qw
        & \qw 
        & \ghost{\mathcal{G}^{(0)}} & \qw
        & \qw 
        & \push{\hspace{0.05cm}\cdots\hspace{0.05cm}} & \qw
        & \qw & \qw
        & \qw 
        & \ghost{\mathcal{G}^{(T-1)}} & \qw
        & \qw
        & \meter{}
    \end{quantikz}
    \caption{Illustration of the iterative interaction between the agent and the environment. The policy modules \(\Pi_\theta^{(t)}\) act sequentially on the action register, while the controlled environment \(\mathcal{E}^{(t)}\) acts on the state and reward registers. After each environment step, \(\mathcal{G}^{(t)}\) stores the reward information in the auxiliary register \(C\), which is measured at the end of the trajectory.}
        \label{fig:interaction}
\end{figure}
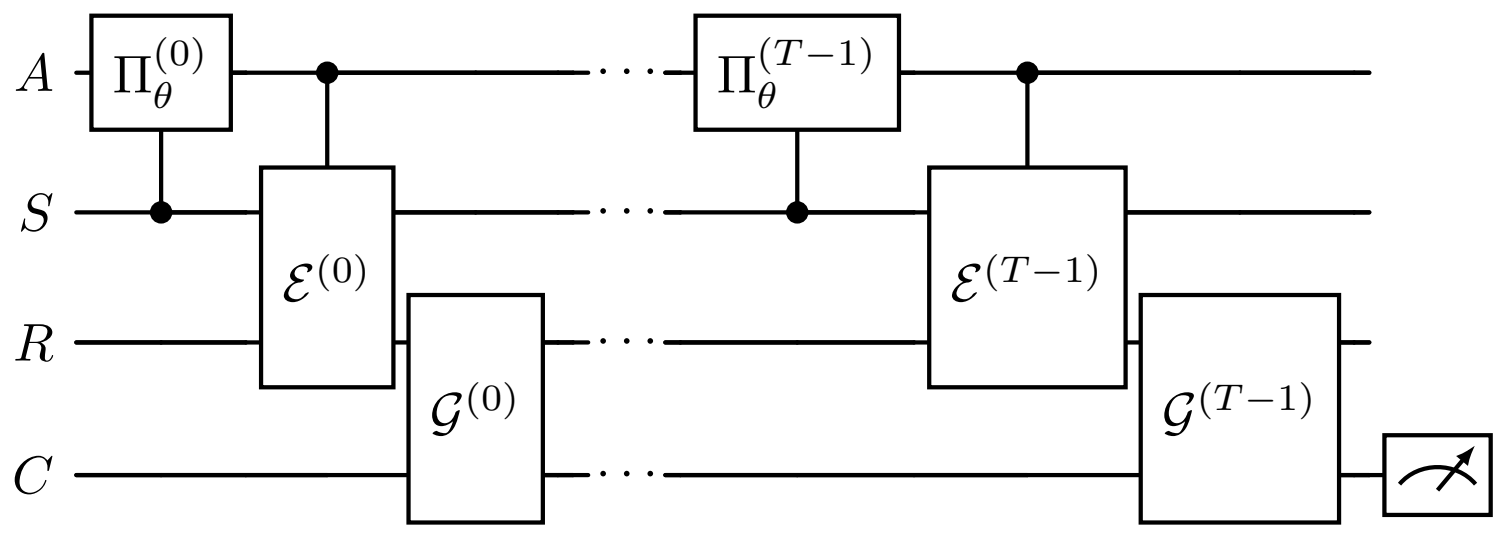

As illustrated in Figure~\ref{fig:interaction}, at time step \(t\), the agent applies a policy CQC \(\Pi_\theta^{(t)}\) that updates the action
register while using the state register only as a control system. Here \(\Pi_{\theta}^{(t)}\) is the quantum analogue of choosing an action according to a policy conditioned on the current state. The important difference is that for choosing the action, the state register is neither measured nor overwritten.

After the action register has been updated, the environment applies an action-controlled CQC. This channel is the quantum counterpart of the classical transition-and-reward rule \(P(s_{t+1},r_t\mid s_t,a_t)\): the action controls how the state and reward registers evolve, but the environment does not directly rewrite the action register. As in classical RL, the agent has no prior knowledge of the environment's CQC.

The reward register \(R\) should be viewed as an instantaneous workspace. To evaluate returns without measuring after each interaction, we introduce a CPTP reward-accumulation map \(\mathcal{G}^{(t)}\) acting jointly on the reward register \(R\) and the auxiliary register \(C\).
One simple choice is to decompose
\(\mathcal H_C=\bigotimes_{t=0}^{T-1}\mathcal H_{C^{(t)}}\), initialize each
\(C^{(t)}\) in a fixed blank state, and set
\[
    \mathcal{G}^{(t)}(\rho) = W^{(t)}\rho \left(W^{(t)}\right)^{\dagger}, \qquad W^{(t)}=\mathrm{SWAP}_{R,C^{(t)}}.
\]
More general CPTP accumulation maps may be used as well, for example to store only task-relevant reward information or to update a smaller accumulator.

We now define one step of the quantum interaction. All maps below are lifted to the full joint Hilbert space by tensoring with identities on registers they do not act on. The one-step interaction channel is
\[
    \Phi_\theta^{(t)} \coloneqq \mathcal G^{(t)} \circ \mathcal{E}^{(t)} \circ \Pi_\theta^{(t)} .
\]
The trajectory state evolves according to
\[
    \rho^{(t+1)} = \Phi_\theta^{(t)}\!\left(\rho^{(t)}\right), \qquad t=0,1,\ldots,T-1.
\]
For trajectories from \(t\) to \(t'\), we write
\[
    \Phi_\theta^{(t':t)} \coloneqq \Phi_\theta^{(t'-1)} \circ \Phi_\theta^{(t'-2)} \circ \cdots \circ \Phi_\theta^{(t)}.
\]

\subsection{Value observable and discounted return}
\label{subsec:quantum-return}

The agent's objective is defined by measuring the accumulated reward information after the interaction has unfolded. Let \(\widehat{R}^{(t)}\) be a bounded Hermitian reward observable acting on \(C^{(t)}\). In the Heisenberg picture, define the value observable from time \(t\) to horizon \(T\) by
\[
    V_\theta^{(T:t)} \coloneqq \left(\Phi_\theta^{(T:t)}\right)^\dagger
    \left(
        \sum_{k=t}^{T-1}
        \gamma^{k-t}\widehat{R}^{(k)}
    \right).
\]
Then for any intermediate trajectory state \(\rho^{(t)}\),
\(
    v_\theta^{(T:t)} \coloneqq
    \expval{
        V_\theta^{(T:t)}
    }_{\rho^{(t)}}
\)
is the quantum analogue of expected discounted return. The finite-horizon performance of the policy is
\[
    v^{(T)}_\theta \coloneqq v^{(T:0)}_\theta.
\]

For the infinite-horizon setting, define the expected discounted return by
\[
    v_\theta \coloneqq \lim_{T\to\infty} v_\theta^{(T)},
\]
whenever the limit exists. 

The policy-optimization problem is therefore to find \(\theta\) that maximizes \(v^{(T)}_\theta\) in the finite-horizon case, or \(v_\theta\) in the infinite-horizon case.

\section{Quantum policy gradient}
\label{sec:qpolicy-gradient}

The previous section defines a fully QRL framework at the level of CQCs. To make this framework operational on gate-based quantum hardware, we instantiate the policy class using CPQCs. 

\begin{figure}
    \centering
    \begin{quantikz}
    \lstick{$X$} & \gate{U_c^\dagger(\theta_1)} & \ctrl{1} & \gate{U_c(\theta_1)} & \qw \\
    \lstick{$Y$} & \qw & \gate{U_t(\theta_2)} & \qw & \qw
    \end{quantikz}
    \caption{CPQC structure. The unitary \(U_c(\theta_1)\) sets the conditional basis, enabling a controlled application of \(U_t(\theta_2)\) on the target register in an arbitrary orthonormal basis. The register \(X\) is the control register and register \(Y\) is the target register.}
    \label{fig:cvqc-intro}
\end{figure}
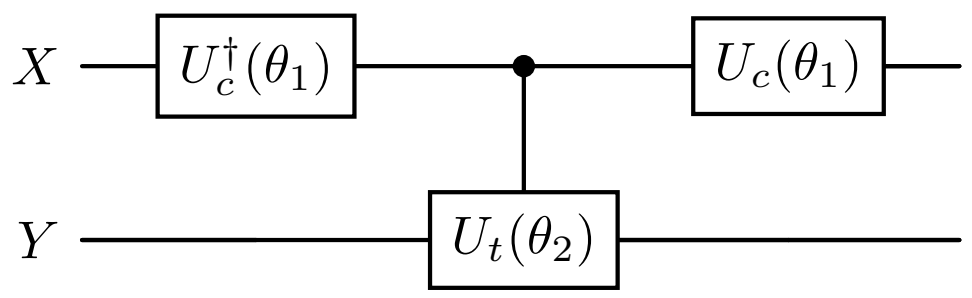

A CPQC is a two-register parameterized CQC with a control register \(X\) and a target register \(Y\). In the policy channel, the control register is the state register \(S\), and the target register is the action register \(A\). The circuit applies a trainable basis-change unitary \(U_c(\theta_1)\) to the control register before and after a standard controlled unitary \(U_t(\theta_2)\) on the target register. Specifically, \(U_c^\dagger(\theta_1)\) maps the learned orthonormal control basis to the computational basis, where the standard controlled application of \(U_t(\theta_2)\) is implemented, and \(U_c(\theta_1)\) maps the computational basis back to the learned control basis; see Figure~\ref{fig:cvqc-intro}. In this way, the circuit learns both the transformation applied to the action register and the basis of the state register. The formal definition of a CPQC and the proof that a CPQC is a valid CQC are deferred to Appendix~\ref{app:cvqc}.

Lemma~\ref{prop:infinite-return-well-defined} below establishes that \(v_\theta\) is well-defined under the bounded-reward and 
discounting assumptions. See Appendix~\ref{app:efficient-psr} for the proof.

\begin{lemma}
\label{prop:infinite-return-well-defined}
For the infinite-horizon objective, assume \(0\leq\gamma<1\) and that the reward
observables are uniformly bounded:
\[
    \|\widehat R^{(t)}\|_{\mathrm{op}}\leq R_{\max},
    \qquad t=0,1,\ldots .
\]
Then the limit \(v_\theta=\lim_{T\to\infty}v_\theta^{(T)}\) exists and is differentiable with respect to each policy parameter \(\theta_k\).
\end{lemma}

As described in the framework, the environment \(\mathcal E\) is a CQC and the reward-accumulation map \(\mathcal G\) is a quantum channel. For the policy, we impose the stationarity assumption used in policy-gradient optimization. During the estimation of the discounted return and its gradient, the policy is fixed: the same CPQC channel $\Pi_\theta$ is applied at every time step of the trajectory. Formally,
\[
    \Pi_\theta^{(t)}=\Pi_\theta,
    \qquad
    t=0,1,\ldots .
\]
The parameter vector $\theta$, whose dimension we denote by \(d\), is therefore shared across time. This means that a single policy is used to generate the full quantum trajectory and to estimate the expected return associated with that policy. After the discounted return gradient is estimated, the parameters are updated, producing a new policy for the next optimization iteration.

For parameter-shift estimates, we define the shared parameter-shift rule for \(\theta_k\), where \(\theta_k\) denotes the target parameter, \(\rho\) the input state, and \(\widehat{R}\) the observable, as
\begin{equation}
    \label{eq:shared-parameter-shift-rule}
    \operatorname{SPSR}_k(\rho, \Pi_\theta, \widehat R)
    \coloneqq
    \begin{cases}
    \dfrac{1}{2}\Big(
        \expval{\widehat R}_{\Pi_{\theta+ \frac{\pi}{2}e_k}(\rho)}
        -
        \expval{\widehat R}_{\Pi_{\theta - \frac{\pi}{2}e_k}(\rho)}
    \Big)
    & \text{if } \theta_k \text{ occurs once},
    \\[1em]
    \begin{aligned}
        &\frac{\sqrt{2}+1}{2\sqrt{2}}\Big(
            \expval{\widehat R}_{\Pi_{\theta+ \frac{\pi}{4}e_k}(\rho)}
            -
            \expval{\widehat R}_{\Pi_{\theta - \frac{\pi}{4}e_k}(\rho)}
        \Big)
        \\
        &\quad
        -
        \frac{\sqrt{2}-1}{2\sqrt{2}}\Big(
            \expval{\widehat R}_{\Pi_{\theta+ \frac{3\pi}{4}e_k}(\rho)}
            -
            \expval{\widehat R}_{\Pi_{\theta - \frac{3\pi}{4}e_k}(\rho)}
        \Big)
    \end{aligned}
    & \text{if } \theta_k \text{ occurs twice},
\end{cases}
\end{equation}
where $e_k$ is the $k$-th standard basis vector. The above equations follow from the standard parameter-shift rule \citep{Schuld_2019,Mitarai_2018} and its generalized form (see Equation~(8) of \citep{wierichs2022general}). 


\subsection{Infinite-horizon optimization}

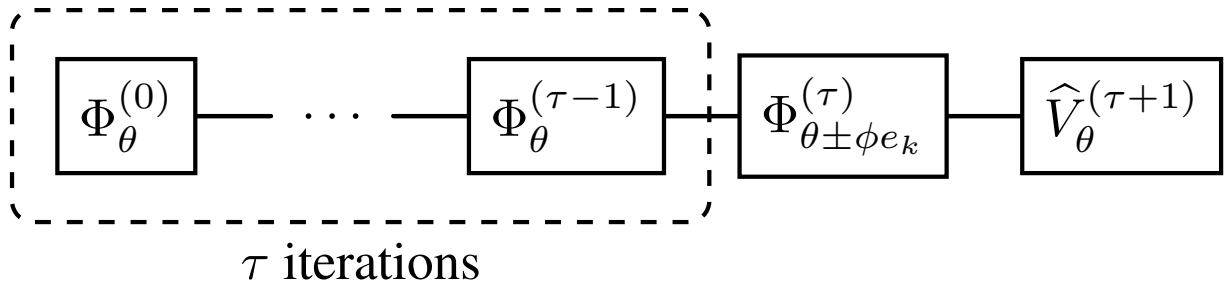
\begin{figure}
    \centering
    \begin{quantikz}
        \gate{\Phi_\theta^{(0)}} \qw \gategroup[1,steps=3,style={dashed, rounded corners, inner sep=5pt}, label style={label position=below, anchor=north, yshift=-0.2cm}]{$\tau$ iterations} & 
        \hspace{0.2cm} \dots \hspace{0.2cm} \qw & 
        \gate{\Phi_\theta^{(\tau-1)}} \qw & 
        \gate{\Phi_{\theta\pm \phi e_k}^{(\tau)}} \qw & 
        \gate{\widehat V_\theta^{(\tau+1)}} \qw
    \end{quantikz}
    \caption{Quantum circuit architecture for gradient estimation at a sampled block index \(\tau\). The sequence consists of \(\tau\) unshifted interaction channels \(\Phi_\theta\), followed by one interaction channel \(\Phi_{\theta}^{(\tau)}\) whose parameters are shifted according to the angles in \(\operatorname{SPSR}_k\), and concludes with the suffix-value estimator \(\widehat V_\theta^{(\tau)}\).}
    \label{fig:gradient-circuit}
\end{figure}

\begin{algorithm}[t]
\caption{Quantum Policy Gradient for Infinite Horizon}
\label{alg:quantum-policy-gradient}
\begin{algorithmic}[1]
\State \textbf{Input:} Initial state $\rho_0$, discount factor $\gamma \in [0,1)$, step-sizes $\{\alpha_n\}_{n\ge 0}$, number of iterations $N$, suffix-value estimator $\widehat V_\theta^{(t)}$, environment channel $\mathcal E$
\State Initialize $\theta \sim \operatorname{Unif}\!\left(\left[0,2\pi\right]^{d}\right)$
\For{$n=0,1,\dots,N-1$}
    \For{$k=1,2,\dots,d$}
        \State Sample \(\tau \sim \operatorname{Geom}(1-\gamma)\), so that \(\Pr(\tau=t)=(1-\gamma)\gamma^t\)
        \State Apply unshifted interactions \(\Phi_{\theta}^{(\tau:0)}\) to obtain \(\rho^{(\tau)}\)
        \State Using fresh circuit executions, estimate \[\widehat g_{k} \leftarrow \operatorname{SPSR}_k \left(
            \rho^{(\tau)}, \Phi_{\theta}^{(\tau)},
            \widehat R^{(\tau)} + \gamma \widehat V_{\theta}^{(\tau+1)}
        \right)/(1-\gamma)\]
    \EndFor
    \State $\widehat g \leftarrow (\widehat g_{1},\dots,\widehat g_{d})$
    \State $\theta \leftarrow \theta + \alpha_n \widehat g$
\EndFor
\State \textbf{return} $\theta$
\end{algorithmic}
\end{algorithm}

Optimizing the infinite-horizon discounted return is challenging because the interaction is not represented by a finite circuit. Thus the standard parameter-shift rule does not apply directly to the infinite-horizon case. In this subsection, we derive an infinite-horizon policy-gradient estimator that avoids this difficulty.

To simplify the presentation of the algorithm, we first make the following
assumption and explain later how it can be implemented. For every time step
\(t\), policy parameter \(\theta\), and input state \(\rho\),
we assume access to a suffix-value estimation procedure
\(\widehat V_\theta^{(t)}\) whose expectation value on \(\rho\) satisfies
\begin{equation}
    \expval{\widehat V_\theta^{(t)}}_\rho
    =
    \lim_{T\to\infty}
    \expval{V_\theta^{(T:t)}}_{\rho},
    \label{eq:unbiased-suffix-value-estimator}
\end{equation}
Thus, on the same input state, the estimator and the exact infinite-horizon
value observable have the same expected measurement.

The intuition behind the algorithm is that earlier policy blocks have a larger effect on the discounted return, because the reward at time \(t\) is weighted by \(\gamma^t\). We therefore sample a zero-based policy-block index \(\tau\) with probabilities
\[
    \operatorname{Pr}(\tau=t)=(1-\gamma)\gamma^t,
    \qquad t=0,1,\ldots,
\]
whose probabilities match these discount weights. As shown in Figure~\ref{fig:gradient-circuit}, the circuit is applied for \(\tau\) unshifted iterations, then the policy block at time \(\tau\) is shifted, and the shifted suffix return is estimated using \(\widehat V_\theta^{(\tau)}\). The shift of the sampled block is the infinite-horizon analogue of the standard parameter-shift rule, and the resulting gradient estimate is used to update the policy parameters. Algorithm~\ref{alg:quantum-policy-gradient} gives a more detailed version of this procedure.

\begin{theorem}[Unbiased gradients in Algorithm~\ref{alg:quantum-policy-gradient}]
\label{thm:infinite-shared-cvqc-gradient}
Suppose that the conditions of Lemma~\ref{prop:infinite-return-well-defined}
hold and that \(\widehat V_\theta^{(t)}\) satisfies the unbiasedness condition
in Equation~\eqref{eq:unbiased-suffix-value-estimator}. Let \(\widehat g_k\) be
the gradient-coordinate estimate produced by
Algorithm~\ref{alg:quantum-policy-gradient}. Then
\[
    \mathbb E\!\left[\widehat g_k\right]
    = \frac{\partial v_\theta}{\partial\theta_k}.
\]
\end{theorem}

The proof begins with a finite horizon and unrolls the gradient recursion to
express the derivative as a discounted sum of one-block
\(\operatorname{SPSR}_k\) terms. Bounded rewards and discounting then permit
passage to the infinite-horizon limit by dominated convergence. Using the
linearity of \(\operatorname{SPSR}_k\), each term is separated into its
immediate-reward and suffix-value contributions, and
Equation~\eqref{eq:unbiased-suffix-value-estimator} allows the limiting exact
suffix observable to be replaced by \(\widehat V_\theta^{(t)}\). Finally,
sampling the block index \(\tau\) geometrically converts the resulting infinite
discounted sum into the expectation of the coordinate estimator used in
Algorithm~\ref{alg:quantum-policy-gradient}, yielding
\(\partial v_\theta/\partial\theta_k=\mathbb E[\widehat g_k]\). The full proof
is in Appendix~\ref{app:inf-proof}.

One way to realize \(\widehat V_\theta^{(t)}\) is to continue the circuit from
an input state at time \(t\) using the unshifted interaction channel for later
steps, and then measure the accumulated discounted reward. In the ideal
infinite-horizon setting, its expected measurement equals
\(\lim_{T\to\infty}\expval{V_\theta^{(T:t)}}_\rho\), as required by
Equation~\eqref{eq:unbiased-suffix-value-estimator}. Such an estimator can be implemented, for example, by randomizing the suffix
horizon according to the discount factor. On hardware, one may implement a truncated estimator by continuing the circuit for \(H\) steps, which introduces a bias that decays
geometrically with the truncation length; see Appendix~\ref{app:truncation-bias}.

\subsection{Convergence analysis}
\label{subsec:convergence-analysis}

The optimization landscape induced by a CPQC policy is generally non-convex, so the appropriate convergence guarantee for Algorithm~\ref{alg:quantum-policy-gradient} is convergence to stationary points rather than to a global maximizer. 

\begin{theorem}[Convergence to stationary] 
    \label{thrm:robbins_condition}
    Assume that the suffix-value estimator \(\widehat V_\theta^{(t)}\) satisfies
    Equation~\eqref{eq:unbiased-suffix-value-estimator}. Suppose also that the step sizes satisfy the Robbins-Monro conditions \citep{robbins1951stochastic}:
    \[
        \alpha_n>0,\qquad
        \sum_{n=0}^{\infty}\alpha_n=\infty,
        \qquad
        \sum_{n=0}^{\infty}\alpha_n^2<\infty,
    \]
    and that \(v_\theta\) has an \(L\)-Lipschitz gradient on the parameter domain.
    Here, \(\theta_n\) and \(\widehat g_n\) denote the policy parameters and the
    gradient estimate at optimization step \(n\), respectively. Then
    \[
        \sum_{n=0}^{\infty}\alpha_n\|\nabla v_{\theta_n}\|_2^2 < \infty
        \qquad \text{almost surely}.
    \]
    If, in addition, the estimator noise is asymptotically exact, in the sense that for every \(\varepsilon>0\),
    \[
        \sum_{n=0}^{\infty}
        \Pr\!\left(
            \|\widehat g_n-\nabla v_{\theta_n}\|_2>\varepsilon
        \right)
        <\infty,
    \]
    then
    \[
        \lim_{n\to\infty}\|\widehat g_n\|_2 = 0
        \qquad \text{almost surely}.
    \]
\end{theorem}


The asymptotic exactness condition on the estimator noise in Theorem~\ref{thrm:robbins_condition} is a natural idealization of increasing estimation accuracy during training. In practice, \(\widehat g_n\) is obtained from a finite number of circuit repetitions and, on hardware, possibly from a finite suffix length. Both sources of error can be driven down by increasing the number of shots and the suffix horizon. Thus the condition is not a new structural assumption on the QRL model; it formalizes the standard requirement that the stochastic gradient becomes sufficiently accurate along the optimization run.

\subsection{Finite-horizon optimization}

The same sampling idea also gives a finite-horizon policy-gradient estimator. For a fixed horizon \(T\), instead of sampling the shifted policy block from an infinite geometric distribution, we sample \(\tau \in\{0,\ldots,T-1\}\) from the truncated geometric distribution
\[
    \Pr(\tau=t)
    =
    \frac{(1-\gamma)\gamma^t}{1-\gamma^T},
    \qquad
    t=0,\ldots,T-1.
\]
This distribution is the finite-horizon analogue of the geometric distribution used in Algorithm~\ref{alg:quantum-policy-gradient}: it gives earlier policy blocks larger probability, while normalizing over the finite set of possible shift locations. The algorithm is therefore almost unchanged. One runs the unshifted interaction up to the sampled index \(\tau\), shifts the \(k\)-th policy parameter in that block by \(\pm\pi/2\), and estimates the remaining finite-horizon suffix return. The only difference from the infinite-horizon estimator is the sampling distribution and the normalization factor.


\begin{theorem}[Unbiased finite-horizon policy-gradient estimator]
\label{cor:finite-horizon-shared-cvqc-gradient}
For the finite-horizon objective \(v_{\theta}^{(T)}\), suppose that the conditions of Lemma~\ref{alg:quantum-policy-gradient} hold and that \(\widehat V_\theta^{(t)}\) satisfies the unbiasedness condition
\[
    \expval{\widehat V_\theta^{(t)}}_\rho
    =
    \expval{V_\theta^{(T:t)}}_{\rho}.
\]
Let \(\widehat g_k\) be
the gradient-coordinate estimate produced by finite version of
Algorithm~\ref{alg:quantum-policy-gradient}. Then
\[
    \mathbb E\!\left[\widehat g_k\right]
    = \frac{\partial v_\theta^{(T)}}{\partial\theta_k}.
\]
\end{theorem}

\begin{proof}
The proof follows the finite-horizon part of the proof of
Theorem~\ref{thm:infinite-shared-cvqc-gradient}. Unrolling the
gradient recursion and applying the shared parameter-shift rule gives
\[
\frac{\partial v_\theta^{(T)}}{\partial\theta_k}
=
\sum_{t=0}^{T-1}\gamma^t
\operatorname{SPSR}_k\!\left(
\rho^{(t)},\Phi_\theta^{(t)},
\widehat R^{(t)}+\gamma\widehat V_\theta^{(t+1)}
\right),
\]
where the suffix-value unbiasedness assumption permits the use of
\(\widehat V_\theta^{(t+1)}\). Sampling \(\tau\) with
\(\Pr(\tau=t)=(1-\gamma)\gamma^t/(1-\gamma^T)\) and multiplying
the sampled term by \((1-\gamma^T)/(1-\gamma)\) turns this sum into
\(\mathbb E[\widehat g_k]\), as claimed.
\end{proof}

The convergence analysis for the finite-horizon estimator follows the same stochastic approximation argument as in Theorem~\ref{thrm:robbins_condition}. In fact, the finite-horizon case is simpler: the objective is defined by a finite circuit, the shifted suffix returns are finite sums, and the second-moment bounds do not require controlling an infinite discounted tail. Thus, under the same Robbins-Monro step-size conditions and asymptotically exact estimator noise, the finite-horizon policy-gradient estimator converges to stationary behavior for \(v_\theta^{(T)}\).

\section{Conclusion and future work}
\label{sec:conclusion}

We introduced a fully QRL framework in which the agent, 
environment, reward, and return accumulation are all represented and processed quantum mechanically, without intermediate measurement. The interaction is structured through separate state, action, reward, and auxiliary registers evolving under CQCs, which preserve the agent-environment separation while allowing the trajectory to remain coherent until final evaluation. This addresses a fundamental limitation of existing hybrid QRL approaches, where repeated measurement collapses entanglement and requires costly state re-preparation at each step.

To make the framework operational on gate-based hardware, we instantiated the policy
class using CPQCs, in which a trainable 
basis-change unitary $U_c(\theta_1)$ allows the policy to learn not only the 
action-generating transformation, but also the basis of the state register in which 
conditioning is performed. This stands in contrast to prior oracular QRL formulations \citep{dunjko2015frameworklearningagentsquantum,Dunjko_2016,Dunjko_2017}, where states and actions are fixed in the computational basis. We derived policy-gradient estimators for both finite- and infinite-horizon settings using a shared-parameter shift rule. In the infinite-horizon case, the key challenge is that the interaction is not representable as a finite circuit; we resolved this by sampling a zero-based policy-block index \(\tau\) with probability \((1-\gamma)\gamma^\tau\), whose discount weights match the contribution of each time step to the discounted return, yielding an unbiased gradient estimator (Theorem~\ref{thm:infinite-shared-cvqc-gradient}). For the finite-horizon case, the gradient decomposes as 
a sum of shifted suffix returns weighted by the external discount factor 
(Theorem~\ref{cor:finite-horizon-shared-cvqc-gradient}).

\paragraph{Limitations and future works.} The present work is theoretical and does not yet evaluate the proposed estimator on quantum-control benchmarks. A full sample-complexity analysis, including shot complexity and the dependence on the discount factor, remains open. The measurement-free formulation also shifts the burden from repeated measurement to coherent circuit depth and auxiliary memory, which may be challenging on noisy near-term hardware. Almost-sure convergence of the resulting stochastic gradient-ascent procedure (Algorithm \ref{alg:quantum-policy-gradient}) to stationary points is established under boundedness conditions and $L$-smoothness of the objective (Theorem~\ref{thrm:robbins_condition}); extending this guarantee beyond these assumptions is an open direction. Finally, a barren plateau analysis for the CPQC policy class is needed, as the trainable control basis introduces additional structure not covered by existing results \citep{McClean2018,Ragone2024}.

\section{Acknowledgments}
This work was partially supported by the NSF Grant CCF-2211423.

\newpage

\bibliographystyle{alpha}
\bibliography{neurips_2026}


\newpage

\appendix

\section{Conditional parameterized quantum circuits}
\label{app:cvqc}

\begin{definition}
    Let \(U_\theta\) denote a CPQC acting on a control register \(X\) and target register \(Y\). 
    The circuit is characterized by a PQC \(U_{c}(\theta_1)\) acting on \(X\), and a controlled PQC \(\mathrm{C}-U_t(\theta_2)\) acting on \(Y\). The CPQC is defined by the unitary
    \[
        U_\theta = \left(
            U_c(\theta_1) \otimes I_Y
        \right) 
            \mathrm{C}-U_t(\theta_2)
        \left(
            U_c^\dagger(\theta_1) \otimes I_Y
        \right).
    \]
    
\end{definition}

\begin{proposition}
    A CPQC is a valid CQC.
\end{proposition}
\begin{proof}
    Let \(\{\ket{i}\}_i\) be the computational basis of the control register \(X\). Then the controlled unitary can be written as
    \[
         \mathrm{C}-U_t(\theta_2) = \sum_i \dyad{i}{i}_X \otimes U_{t,i}(\theta_2),
    \]
    where \(U_{t,i}\) is the PQC applied to the target register \(Y\) when the control register \(X\) is \(\ket{i}\).

    Define the projective resolution
    \[
        P_{i}(\theta_1) \coloneq U_{c}(\theta_1) \dyad{i}{i}_X U^\dagger_c(\theta_1).
    \]
    Note that \(P_i(\theta_1)= P^\dagger_i(\theta_1)\), \(P_i(\theta_1) P_j(\theta_1) = \delta_{ij}P_i(\theta_1)\), and
    \[
        \sum_i P_i(\theta_1) = U_c(\theta_1) \left(\sum_i \dyad{i}{i}_X\right) U^\dagger_c(\theta_1) = I_X.
    \]
\end{proof}

\section{Proofs for infinite-horizon optimization}
\label{app:efficient-psr}

\subsection{Proof of Lemma~\ref{prop:infinite-return-well-defined}: existence of the limit}

We first prove the existence of the limit. To do so, we show that \(v_\theta^{(T:t)}\) is a Cauchy sequence; that is, for every \(\epsilon > 0\), there exists an integer \(N\) such that for all integers \(m > n > N\),
\[|v_\theta^{(m:t)} - v_\theta^{(n:t)}| < \epsilon.\]

\begin{proposition}
    \label{prop:summability-of-v}
    Suppose that, for every \(i\), each later interaction channel
    \(\Phi_\theta^{(j)}\), \(j>i\), acts trivially on the reward sub-register
    \(C^{(i)}\).  Then, for any integers \(m>t\),
    \[
        v_\theta^{(m:t)}
        = \sum_{i=t}^{m-1}\gamma^{i-t}
        \expval{\widehat R^{(i)}}_{\rho^{(i+1)}}.
    \]
\end{proposition}

\begin{proof}
    Based on the definition of \(v_\theta^{(m:t)}\),
    \begin{align*}
        v_\theta^{(m:t)}
        &=\Tr\!\left[
            \left(\Phi_\theta^{(m:t)}\right)^\dagger
            \left(\sum_{i=t}^{m-1}\gamma^{i-t}\widehat R^{(i)}\right)
            \rho^{(t)}
          \right]\\
        &=\sum_{i=t}^{m-1}\gamma^{i-t}
          \Tr\!\left[
            \widehat R^{(i)}
            \Phi_\theta^{(m:t)}\!\left(\rho^{(t)}\right)
          \right].
    \end{align*}

    Because \(\widehat R^{(i)}\) acts only on \(C^{(i)}\), while every
    channel in \(\Phi_\theta^{(m:i+1)}\) acts trivially on that
    sub-register,
    \[
        \left(\Phi_\theta^{(m:i+1)}\right)^\dagger
        \!\left(\widehat R^{(i)}\right)=\widehat R^{(i)}.
    \]
    Therefore we have
    \begin{align*}
        \Tr\!\left[
          \widehat R^{(i)}
          \Phi_\theta^{(m:t)}\!\left(\rho^{(t)}\right)
        \right]
        &=\Tr\!\left[
          \widehat R^{(i)}
          \Phi_\theta^{(i+1:t)}\!\left(\rho^{(t)}\right)
        \right]\\
        &=\Tr\!\left[\widehat R^{(i)}\rho^{(i+1)}\right]
         =\expval{\widehat R^{(i)}}_{\rho^{(i+1)}}.
    \end{align*}
    Substitution into the preceding sum proves the claim.
\end{proof}

\begin{corollary}
    For any integers \(m > n > t\), the finite-horizon value can be written as
    \[
        v_\theta^{(m:t)} = v_\theta^{(n:t)} + \gamma^{n-t} v_\theta^{(m:n)}.
    \]
\end{corollary}
\begin{proof}
    Split the sum in Proposition~\ref{prop:summability-of-v} at \(n\),
    \begin{align*}
        v_\theta^{(m:t)}
        &=\sum_{i=t}^{n-1}\gamma^{i-t}
          \expval{\widehat R^{(i)}}_{\rho^{(i+1)}}
          +\sum_{i=n}^{m-1}\gamma^{i-t}
          \expval{\widehat R^{(i)}}_{\rho^{(i+1)}}\\
        &=v_\theta^{(n:t)}
          +\gamma^{n-t}\sum_{i=n}^{m-1}\gamma^{i-n}
          \expval{\widehat R^{(i)}}_{\rho^{(i+1)}}\\
        &=v_\theta^{(n:t)}+\gamma^{n-t}v_\theta^{(m:n)}.
    \end{align*}
\end{proof}

Hence, for any \(m>n>t\),
\begin{align*}
    \left|v_\theta^{(m:t)}-v_\theta^{(n:t)}\right|
    &=\gamma^{n-t}\left|v_\theta^{(m:n)}\right|\\
    &=\gamma^{n-t}\left|
      \sum_{k=n}^{m-1}\gamma^{k-n}
      \expval{\widehat R^{(k)}}_{\rho^{(k+1)}}\right|.
\end{align*}

Since, by assumption, the observables \(\widehat R^{(k)}\) are uniformly bounded by some constant \(R_\mathrm{max}\), we can apply the triangle inequality to get
\begin{align*}
     \left|v_\theta^{(m:t)}-v_\theta^{(n:t)}\right|
     &\leq \gamma^{n-t}\sum_{k=n}^{m-1}\gamma^{k-n}
       \left|\expval{\widehat R^{(k)}}_{\rho^{(k+1)}}\right|\\
     &\leq \gamma^{n-t}R_\mathrm{max}
       \sum_{j=0}^{m-n-1}\gamma^j\\
     &\leq \gamma^{n-t}R_\mathrm{max} \frac{1-\gamma^{m-n}}{1-\gamma}\\
     &\leq R_\mathrm{max}\frac{\gamma^{n-t}}{1-\gamma}.
\end{align*}
The right-hand side can be made smaller than any \(\epsilon\) by making \(n-t\) sufficiently large. Thus \(v_\theta^{(T:t)}\) is a Cauchy sequence, so the limit exists and is finite. 

\subsection{Proof of Lemma~\ref{prop:infinite-return-well-defined}: differentiability}

Next, we establish the differentiability of \(v^{(t)}_\theta = \lim_{T \to \infty} v^{(T:t)}_\theta\). 
We first state the basic shared-parameter rule in a form that is not specific to the QRL framework. It applies to any finite circuit obtained by repeating the same parameterized VQC block.

\begin{lemma}[Shared-parameter shift rule]
\label{prop:shared-parameter-repeated-blocks}

Let \(J_\theta\) be the expectation of an arbitrary observable \(M\) on a parameterized circuit. Suppose \(\theta_k\) appears in the circuit at the set of occurrences \(\mathcal{S}_k = \{j= 1, \ldots, m_k\}\), each of which is a gate of the form \(e^{-i\theta_kP_k/2}\) where \(P_k\) has eigenvalues \(\pm1\). Define \(J_{\theta, \pm k, j}\) as the value of the circuit where only the \(j\)-th occurrence of \(\theta_k\) is shifted by \(\pm\pi/2\), and all other gates are unchanged. Then
\[
    \frac{\partial J_\theta}{\partial \theta_k} = \frac{1}{2}\sum_{j\in\mathcal{S}_k} \left(
        J_{\theta, +k, j} - J_{\theta, -k, j}
    \right).
\]

\end{lemma}
\begin{proof}
By the multivariate chain rule, the derivative with respect to the shared parameter is the sum of the partial derivatives with respect to each occurrence. Each partial derivative involves only one gate occurrence, so the standard parameter-shift rule applies. Summing over the repeated blocks gives the result.
\end{proof}

We also rely on the following standard theorem from real analysis regarding the uniform convergence of derivatives; see Theorem 7.17 of \citep{rudin2021principles}.

\begin{lemma}
    \label{lemma:inf-derivative}
     Let \(f_n\) be defined and differentiable on a closed interval \(I\), and let the derivatives \(D f_n\) be uniformly convergent on \(I\). Then, if \(f_n\) converges at some point of \(I\), it converges uniformly on all of \(I\) to a differentiable limit, and 
     \[ D\!\left(\lim_{n\to\infty} f_n\right) = \lim_{n\to\infty} D f_n. \]
\end{lemma}

For any component \(k\) of \(\theta\), we apply this theorem to \(f_n(\theta_k) \coloneqq v^{(n:t)}_\theta\) for all \(\theta_k \in [-\pi, \pi]\). We have shown above that \(v^{(T:t)}_\theta\) is uniformly convergent for any \(\theta\). 
In addition, \(v^{(T:t)}_\theta\) is differentiable for any \(T\), since the circuit consists of smooth exponential generators (Pauli rotations). To prove the uniform convergence of \(\frac{\partial}{\partial\theta_{k}}v_{\theta}^{(T:t)}\), we show that it forms a uniformly Cauchy sequence. The argument is similar to the proof of uniform convergence of \(v^{(T:t)}_\theta\) in the first part of the proof. 

For any integers $m > n > t$,
\begin{align*}
\left| \frac{\partial}{\partial\theta_{k}}v_{\theta}^{(m:t)} - \frac{\partial}{\partial\theta_{k}}v_{\theta}^{(n:t)} \right| &= \left| \frac{\partial}{\partial\theta_{k}} \left( \sum_{l=t}^{m-1} \gamma^{l-t}\langle \widehat R^{(l)}\rangle_{\rho^{(l+1)}} \right) - \frac{\partial}{\partial\theta_{k}} \left( \sum_{l=t}^{n-1} \gamma^{l-t}\langle \widehat R^{(l)}\rangle_{\rho^{(l+1)}} \right) \right| \\
&= \left| \sum_{l=n}^{m-1} \gamma^{l-t} \frac{\partial}{\partial\theta_{k}}\langle \widehat R^{(l)}\rangle_{\rho^{(l+1)}} \right| \\
&\leq \sum_{l=n}^{m-1} \gamma^{l-t} \left| \frac{\partial}{\partial\theta_{k}}\langle \widehat R^{(l)}\rangle_{\rho^{(l+1)}} \right|.
\end{align*}

Let \(\mathcal S_k\) be the occurrences of \(\theta_k\) until step \(l\). Applying
Lemma~\ref{prop:shared-parameter-repeated-blocks}, we can therefore write
\[
\frac{\partial}{\partial\theta_k}
\left\langle \widehat R^{(l)}\right\rangle_{\rho^{(l+1)}}
=
\frac{1}{2}\sum_{j\in \mathcal S_k}
    \left\langle \widehat R^{(l)}\right\rangle_{\rho^{(l+1)}_{+j}} - 
    \left\langle \widehat R^{(l)}\right\rangle_{\rho^{(l+1)}_{-j}},
\]
where \(\rho_{\pm j}^{(l+1)}\) is the mixed state obtained by shifting only the
\(j\)-th occurrence of \(\theta_k\) by \(\pm\pi/2\). The bounded-reward assumption and the triangle inequality consequently imply
\begin{align*}
\left| \frac{\partial}{\partial\theta_{k}}
\left\langle \widehat R^{(l)}\right\rangle_{\rho^{(l+1)}} \right|
&\leq \frac{1}{2}\sum_{j\in\mathcal S_k}
\left|\left\langle \widehat R^{(l)}\right\rangle_{\rho^{(l+1)}_{+j}}\right|+
\left|
\left\langle \widehat R^{(l)}\right\rangle_{\rho^{(l+1)}_{-j}}
\right|\\
&\leq \left|\mathcal S_k\right|R_{\max}.
\end{align*}

Based on the structure of the CPQC, \(\left|\mathcal S_k\right| \leq 2(l + 1)\). Substituting this bound back into the inequality yields

\begin{align*}
 \left| \frac{\partial}{\partial\theta_{k}}v_{\theta}^{(m:t)} - \frac{\partial}{\partial\theta_{k}}v_{\theta}^{(n:t)} \right| &\le \sum_{l=n}^{m-1} 2\gamma^{l-t} (l+1) R_{\max} \\
&= 2 R_{\max} \gamma^{n-t} \sum_{j=0}^{m-n-1} \gamma^{j} (n + j + 1) \\
&\leq 2 R_{\max} \gamma^{n-t} \left( (n+1)\sum_{j=0}^\infty \gamma^j + \sum_{j=0}^\infty \gamma^j j \right) \\
&= 2 R_{\max} \gamma^{n-t} \left( \frac{n+1}{1-\gamma} + \frac{\gamma}{(1-\gamma)^2}\right) = O(n\gamma^{n-t}),
\end{align*}
which converges to 0 as \(n-t\rightarrow \infty\) because \(\gamma < 1\).

Thus, Lemma~\ref{lemma:inf-derivative} guarantees that $v^{(t)}_\theta = \lim_{T \to \infty} v^{(T:t)}_\theta$ is differentiable and
\[\frac{\partial}{\partial \theta_k} v^{(t)}_\theta = \lim_{T \to \infty} \frac{\partial}{\partial \theta_k} v^{(T:t)}_\theta.\]

\subsection{Proof of Theorem~\ref{thm:infinite-shared-cvqc-gradient}}
\label{app:inf-proof}
We now prove the unbiased infinite-horizon policy-gradient estimator. Define the quantum action-value channel \(Q_\theta^{(T:t)}\) by
\[
    Q_\theta^{(T:t)} \coloneqq (\widehat R^{(t)} + \gamma V^{(T:t+1)}_{\theta}) \circ \mathcal{G}^{(t)} \circ \mathcal E^{(t)}.
\]
Then we can rewrite $v^{(T:t)}_\theta$ as
\[
    v^{(T: t)}_\theta = \langle Q^{(T:t)}_\theta \rangle_{\Pi_{\theta}(\rho^{(t)})}.
\]
Differentiating with respect to the parameter $\theta_k$ yields two terms via the product rule
\[
    \frac{\partial v^{(T:t)}_\theta}{\partial \theta_k} = \left\langle \frac{\partial  Q^{(T:t)}_\theta}{\partial \theta_k} \right\rangle_{\Pi_{\theta}(\rho^{(t)})} + \langle  Q^{(T:t)}_\theta \rangle_{\frac{\partial \Pi_{\theta}}{\partial \theta_k}(\rho^{(t)})}.
\]
Because the reward observable $\widehat R^{(t)}$, the environment map $\mathcal E^{(t)}$, and the reward-accumulation map $\mathcal{G}^{(t)}$ are independent of $\theta$, the derivative of $Q^{(T,t)}_\theta$ simplifies to
\[
    \frac{\partial Q^{(T:t)}_\theta}{\partial \theta_k} = \gamma \frac{\partial V^{(T:t+1)}_{\theta}}{\partial \theta_k} \circ \mathcal{G}^{(t)} \circ \mathcal E^{(t)}.
\]
Substituting this back into the derivative of the expected discounted return
\[
    \frac{\partial v^{(T:t)}_\theta}{\partial \theta_k} = \gamma \left\langle \frac{\partial V^{(T:t+1)}_{\theta}}{\partial \theta_k} \circ \mathcal{G}^{(t)} \circ \mathcal E^{(t)} \right\rangle_{\Pi_{\theta}(\rho^{(t)})} + \langle  Q^{(T:t)}_\theta \rangle_{\frac{\partial \Pi_{\theta}}{\partial \theta_k}(\rho^{(t)})}.
\]
Using the definition of the interaction channel $\Phi^{(t)}_\theta = \mathcal{G}^{(t)} \circ \mathcal E^{(t)} \circ \Pi_{\theta}$, the first term can be rewritten as
\[
    \frac{\partial v^{(T:t)}_\theta}{\partial \theta_k} = \gamma \left\langle \frac{\partial V^{(T:t+1)}_\theta}{\partial \theta_k} \right\rangle_{\rho^{(t+1)}} + \langle  Q^{(T:t)}_\theta \rangle_{\frac{\partial \Pi_{\theta}}{\partial \theta_k}(\rho^{(t)})}.
\]

Recognizing that the first term is exactly $\gamma \frac{\partial v^{(T:t+1)}_\theta}{\partial \theta_k}$, we obtain the recursive relationship
\[
     \frac{\partial v^{(T:t)}_\theta}{\partial \theta_k} = \gamma \frac{\partial v^{(T:t+1)}_\theta}{\partial \theta_k} + \langle  Q^{(T:t)}_\theta \rangle_{\frac{\partial \Pi_{\theta}}{\partial \theta_k}(\rho^{(t)})}.
\]
By unrolling this recursion, we arrive at the following equation
\[
    \frac{\partial v^{(T:t)}_\theta}{\partial \theta_k} = \sum_{t'=t}^{T-1} \gamma^{t'-t} \langle  Q^{(T:t')}_\theta \rangle_{\frac{\partial \Pi_{\theta}}{\partial \theta_k}(\rho^{(t')})}.
\]
Applying the shared parameter-shift rule \(\operatorname{SPSR}_k\) from
Equation~\eqref{eq:shared-parameter-shift-rule} gives
\[
    \frac{\partial v^{(T:t)}_\theta}{\partial \theta_k}
    = \sum_{t'=t}^{T-1}\gamma^{t'-t}
    \operatorname{SPSR}_k\!\left(
        \rho^{(t')},\Pi_\theta,
        Q_\theta^{(T:t')}
    \right).
\]
This step uses both the standard parameter-shift rule and its generalized form
for parameters that occur more than once in a policy block
\citep{Schuld_2019,Mitarai_2018,wierichs2022general}.

We next pass to the infinite-horizon objective. By
Lemma~\ref{prop:infinite-return-well-defined},
\begin{align*}
    \frac{\partial v_\theta^{(t)}}{\partial \theta_k}
    &= \lim_{T\to\infty}
    \frac{\partial v_\theta^{(T:t)}}{\partial \theta_k}\\
    &= \lim_{T\to\infty}
    \sum_{t'=t}^{T-1}\gamma^{t'-t}
    \operatorname{SPSR}_k\!\left(
        \rho^{(t')},\Pi_\theta,
        Q_\theta^{(T:t')}
    \right).
\end{align*}
Using an indicator function, the last expression can be written as
\[
    \frac{\partial v_\theta^{(t)}}{\partial \theta_k}
    = \lim_{T\to\infty}\sum_{t'=t}^{\infty}
    \mathbf 1_{\{t'<T\}}\gamma^{t'-t}
    \operatorname{SPSR}_k\!\left(
        \rho^{(t')},\Pi_\theta,
        Q_\theta^{(T:t')}
    \right).
\]

Since the reward observable is bounded by \(R_\mathrm{max}\), Lebesgue's
dominated convergence theorem (see Theorem 1.34 of
\cite{rudin1987real}) gives
\begin{align*}
    \frac{\partial v_\theta^{(t)}}{\partial \theta_k}
    &= \sum_{t'=t}^{\infty}\gamma^{t'-t}
    \lim_{T\to\infty}\mathbf 1_{\{t'<T\}}
    \operatorname{SPSR}_k\!\left(
        \rho^{(t')},\Pi_\theta,
        Q_\theta^{(T:t')}
    \right)\\
    &= \sum_{t'=t}^{\infty}\gamma^{t'-t}
    \lim_{T\to\infty}
    \operatorname{SPSR}_k\!\left(
        \rho^{(t')},\Pi_\theta,
        Q_\theta^{(T:t')}
    \right)\\
    &= \sum_{t'=t}^{\infty}\gamma^{t'-t}
    \lim_{T\to\infty}
    \operatorname{SPSR}_k\!\left(
        \rho^{(t')},\Phi_\theta^{(t')},
        \widehat R^{(t')} + \gamma V_\theta^{(T:t'+1)}
    \right)
\end{align*}

Since \(\operatorname{SPSR}_k\) is linear in its observable argument, we can
split the last equation to obtain
\begin{align*}
    \frac{\partial v_\theta^{(t)}}{\partial \theta_k}
    &= \sum_{t'=t}^{\infty}\gamma^{t'-t}
    \Bigg[
        \operatorname{SPSR}_k\!\left(
            \rho^{(t')},\Phi_\theta^{(t')},\widehat R^{(t')}
        \right) \\
    &\hspace{4.5cm}
        +\gamma\lim_{T\to\infty}
        \operatorname{SPSR}_k\!\left(
            \rho^{(t')},\Phi_\theta^{(t')},
            V_\theta^{(T:t'+1)}
        \right)
    \Bigg].
\end{align*}

Equation~\eqref{eq:unbiased-suffix-value-estimator} holds for every input
state, including each shifted output state appearing in
\(\operatorname{SPSR}_k\). Therefore,
\[
    \lim_{T\to\infty}
    \operatorname{SPSR}_k\!\left(
        \rho^{(t')},\Phi_\theta^{(t')},V_\theta^{(T:t'+1)}
    \right)
    =
    \operatorname{SPSR}_k\!\left(
        \rho^{(t')},\Phi_\theta^{(t')},\widehat V_\theta^{(t'+1)}
    \right).
\]
Substituting this equality and using linearity once more gives
\[
    \frac{\partial v_\theta^{(t)}}{\partial \theta_k}
    = \sum_{t'=t}^{\infty}\gamma^{t'-t}
    \operatorname{SPSR}_k\!\left(
        \rho^{(t')},\Phi_\theta^{(t')},
        \widehat R^{(t')}+\gamma\widehat V_\theta^{(t'+1)}
    \right).
\]

Let \(\tau\) be a random variable on \(\{t,t+1,\ldots\}\) with probability distribution
\[
    \Pr(\tau=t')=(1-\gamma)\gamma^{t'-t}.
\]
By applying the normalization factor \(1-\gamma\), the above summation can be written as 
\[
    \frac{\partial v_\theta^{(t)}}{\partial \theta_k}
    = \frac{1}{1-\gamma}\,
    \mathbb E_\tau\!\left[
        \operatorname{SPSR}_k\!\left(
            \rho^{(\tau)},\Phi_\theta^{(\tau)},
            \widehat R^{(\tau)}+\gamma\widehat V_\theta^{(\tau+1)}
        \right)
    \right].
\]
The quantity inside the expectation, together with the factor
\(1/(1-\gamma)\), is the coordinate estimator \(\widehat g_k\) in
Algorithm~\ref{alg:quantum-policy-gradient}. Hence,
\[
    \mathbb E[\widehat g_k]
    =
    \frac{\partial v_\theta^{(t)}}{\partial \theta_k}.
\]
Setting \(t=0\) gives the statement of Theorem~\ref{thm:infinite-shared-cvqc-gradient}.

\section{Convergence analysis}
\label{app:convergence}
Let \(\mathcal F_n\) denote the filtration generated by all randomness up to iteration \(n\). By
Theorem~\ref{thm:infinite-shared-cvqc-gradient}, the geometric parameter-shift estimator is unbiased for each coordinate:
\[
    \mathbb E[\widehat g_{n,k}\mid \mathcal F_n] =
    \frac{\partial v_{\theta_n}}{\partial \theta_k}.
\]
Therefore
\[
    \mathbb E[\widehat g_n\mid \mathcal F_n]=\nabla v_{\theta_n}.
\]

The estimator also has a bounded second moment. The observable
\(\widehat R^{(t)}+\gamma\widehat V_\theta^{(t+1)}\) has norm at most
\(R_{\max}/(1-\gamma)\), and the absolute coefficients in
\(\operatorname{SPSR}_k\) sum to at most \(2\). Hence
\[
    \left|\operatorname{SPSR}_k\!\left(
        \rho^{(t)},\Phi_\theta^{(t)},
        \widehat R^{(t)}+\gamma\widehat V_\theta^{(t+1)}
    \right)\right|
    \le \frac{2R_{\max}}{1-\gamma}.
\]
Using the definition of \(\widehat g_{n,k}\),
\[
    |\widehat g_{n,k}|
    \le
    \frac{2R_{\max}}{(1-\gamma)^2},
\]
and therefore
\[
    \|\widehat g_n\|_2^2
    \le
    \frac{4dR_{\max}^2}{(1-\gamma)^4}
    =: C_g .
\]

Since \(v_\theta\) has an \(L\)-Lipschitz gradient, \(v_\theta\) is \(L\)-smooth, and the ascent lemma \citep[Theorem~2.1.5]{nesterov2018lectures} gives
\[
    v_{\theta_{n+1}}
    \ge
    v_{\theta_n}
    + \alpha_n\langle \nabla v_{\theta_n},\widehat g_n\rangle
    - \frac{L}{2}\alpha_n^2\|\widehat g_n\|_2^2 .
\]
Taking conditional expectation with respect to \(\mathcal F_n\) and using the unbiasedness and
boundedness of \(\widehat g_n\), we obtain
\[
    \mathbb E[v_{\theta_{n+1}}\mid \mathcal F_n]
    \ge
    v_{\theta_n}
    + \alpha_n\|\nabla v_{\theta_n}\|_2^2
    - \frac{L C_g}{2}\alpha_n^2 .
\]

Moreover, bounded rewards imply that the objective is uniformly bounded above:
\[
    |v_{\theta}|\le \frac{R_{\max}}{1-\gamma}.
\]
Thus the Robbins-Siegmund theorem \citep{robbins1971convergence} implies that \(v_{\theta_n}\) converges
almost surely and that
\[
    \sum_{n=0}^{\infty}\alpha_n\|\nabla v_{\theta_n}\|_2^2 < \infty
    \qquad \text{almost surely}.
\]
Because the Robbins-Monro step-size condition gives \(\sum_n \alpha_n=\infty\) \citep{robbins1951stochastic}, this already implies
\[
    \liminf_{n\to\infty}\|\nabla v_{\theta_n}\|_2=0 .
\]

To strengthen this to convergence of the full gradient norm, note that the estimator \(\widehat g_n\) is uniformly
bounded, so
\[
    \|\theta_{n+1}-\theta_n\|_2
    \le
    \alpha_n\sqrt{C_g}.
\]
Since \(\nabla v_\theta\) is \(L\)-Lipschitz, the gradient norm cannot jump by more than
\(L\alpha_n\sqrt{C_g}\) in one step. If \(\|\nabla v_{\theta_n}\|_2\) exceeded some \(\varepsilon>0\) infinitely often, then, because its liminf is zero, there would be infinitely many disjoint excursions from at least \(\varepsilon\) down to at most \(\varepsilon/2\). Each such excursion would require a positive amount of cumulative step size and would contribute a positive
amount to \(\sum_n \alpha_n\|\nabla v_{\theta_n}\|_2^2\), contradicting the finite summability established above.
Hence
\[
    \lim_{n\to\infty}\|\nabla v_{\theta_n}\|_2=0
    \qquad \text{almost surely}.
\]

It remains to pass from convergence of the true gradient to convergence of the stochastic estimator used in the algorithm. Write the estimation error as
\[
    \xi_n \coloneqq \widehat g_n-\nabla v_{\theta_n}.
\]
Under the asymptotic exactness assumption, the Borel-Cantelli lemma \citep[Theorem~4.3]{billingsley1995probability} implies that
\[
    \|\xi_n\|_2\to 0
    \qquad \text{almost surely}.
\]
Therefore
\[
    \|\widehat g_n\|_2
    \le
    \|\nabla v_{\theta_n}\|_2+\|\xi_n\|_2
    \to 0
    \qquad \text{almost surely}.
\]

\section{Truncation bias for suffix-return estimates}
\label{app:truncation-bias}

\begin{proposition}
\label{prop:truncation-bias}
    For finite \(H\), the truncation bias in each coordinate of the gradient estimator is bounded by \(\frac{\gamma^H}{(1-\gamma)^2}R_\mathrm{max}\).
\end{proposition}

\begin{proof}
    Fix a shifted return and compare the ideal infinite-horizon suffix return with the version that keeps only the next \(H\) interaction steps. The two quantities differ only in the rewards that occur after the truncation point. Since the reward observables are uniformly bounded by \(R_\mathrm{max}\), the absolute value of the omitted discounted tail is bounded by
    \[
        \sum_{j=H}^{\infty}\gamma^j R_\mathrm{max}
        =
        \frac{\gamma^H}{1-\gamma}R_\mathrm{max}.
    \]
    In the gradient estimator, each shifted return is multiplied by the geometric normalization factor \(1/(1-\gamma)\). Therefore the induced bias in each shifted-return contribution to the estimator is at most
    \[
        \frac{1}{1-\gamma}
        \cdot
        \frac{\gamma^H}{1-\gamma}R_\mathrm{max}
        =
        \frac{\gamma^H}{(1-\gamma)^2}R_\mathrm{max}.
    \]
\end{proof}

\section{Conditional Quantum Channels}

A conditional quantum channel is not the same as first projecting the control register and then applying a channel conditioned on the projection outcome. In particular, it is different from
\[
    \rho_{XY}
    \mapsto
    \sum_i (\mathcal P_i\otimes \mathcal N_i)(\rho_{XY}),
    \qquad
    \mathcal P_i(\omega)=P_i\omega P_i,
\]
where \(\mathcal N_i(\sigma)=\sum_x F_{i,x}\sigma F_{i,x}^\dagger\). This map removes all coherence between different control subspaces. 

The CNOT gate gives a simple example. Let \(X\) be a one-qubit control register and \(Y\) a one-qubit target register. Take
\[
    P_0=\dyad{0}{0},
    \qquad
    P_1=\dyad{1}{1},
\]
and use a single Kraus label with
\[
    F_0=I_Y,
    \qquad
    F_1=X_Y,
\]
where \(X_Y\) denotes the Pauli-\(X\) operator on the target register. Then the CQC has the single Kraus operator
\[
    K=P_0\otimes I_Y+P_1\otimes X_Y,
\]
so
\[
    \mathcal N_{\mathrm{CNOT}}(\rho)=K\rho K^\dagger .
\]
If the input is \((\alpha\ket{0}+\beta\ket{1})\otimes\ket{0}\), this channel produces
\[
    \alpha\ket{00}+\beta\ket{11},
\]
which preserves coherence between the two control branches. The map above would instead produce the mixed state
\[
    |\alpha|^2\dyad{00}{00}+|\beta|^2\dyad{11}{11}.
\]


\newpage
\section*{NeurIPS Paper Checklist}

\begin{enumerate}

\item {\bf Claims}
    \item[] Question: Do the main claims made in the abstract and introduction accurately reflect the paper's contributions and scope?
    \item[] Answer: \answerYes{} 
    \item[] Justification: The abstract and introduction accurately state the four contributions: the QRL framework, CPQC policy, finite/infinite-horizon gradient estimator, and convergence result.

\item {\bf Limitations}
    \item[] Question: Does the paper discuss the limitations of the work performed by the authors?
    \item[] Answer: \answerYes{}
    \item[] Justification: Limitations are discussed in Section 5, including non-convexity, absence of barren plateau analysis, and empirical validation.

\item {\bf Theory assumptions and proofs}
    \item[] Question: For each theoretical result, does the paper provide the full set of assumptions and a complete (and correct) proof?
    \item[] Answer: \answerYes{} 
    \item[] Justification: All theorems include full proofs in Appendices; assumptions are stated in each theorem.

    \item {\bf Experimental result reproducibility}
    \item[] Question: Does the paper fully disclose all the information needed to reproduce the main experimental results of the paper to the extent that it affects the main claims and/or conclusions of the paper (regardless of whether the code and data are provided or not)?
    \item[] Answer: \answerNA{} 
    \item[] Justification: The paper contains no experiments.

\item {\bf Open access to data and code}
    \item[] Question: Does the paper provide open access to the data and code, with sufficient instructions to faithfully reproduce the main experimental results, as described in supplemental material?
    \item[] Answer: \answerNA{} 
    \item[] Justification: The paper contains no experiments.

\item {\bf Experimental setting/details}
    \item[] Question: Does the paper specify all the training and test details (e.g., data splits, hyperparameters, how they were chosen, type of optimizer) necessary to understand the results?
    \item[] Answer: \answerNA{} 
    \item[] Justification: The paper contains no experiments.

\item {\bf Experiment statistical significance}
    \item[] Question: Does the paper report error bars suitably and correctly defined or other appropriate information about the statistical significance of the experiments?
    \item[] Answer: \answerNA{} 
    \item[] Justification: The paper contains no experiments.

\item {\bf Experiments compute resources}
    \item[] Question: For each experiment, does the paper provide sufficient information on the computer resources (type of compute workers, memory, time of execution) needed to reproduce the experiments?
    \item[] Answer: \answerNA{} 
    \item[] Justification: The paper contains no experiments.
    
\item {\bf Code of ethics}
    \item[] Question: Does the research conducted in the paper conform, in every respect, with the NeurIPS Code of Ethics \url{https://neurips.cc/public/EthicsGuidelines}?
    \item[] Answer: \answerYes{} 
    \item[] Justification: The research is foundational theoretical work with no ethical risks.

\item {\bf Broader impacts}
    \item[] Question: Does the paper discuss both potential positive societal impacts and negative societal impacts of the work performed?
    \item[] Answer: \answerYes{} 
    \item[] Justification: Positive impacts include advancing quantum control and coherent feedback; no direct negative societal impacts are identified.
    
\item {\bf Safeguards}
    \item[] Question: Does the paper describe safeguards that have been put in place for responsible release of data or models that have a high risk for misuse (e.g., pre-trained language models, image generators, or scraped datasets)?
    \item[] Answer: \answerNA{} 
    \item[] Justification: The paper poses no risks of misuse.

\item {\bf Licenses for existing assets}
    \item[] Question: Are the creators or original owners of assets (e.g., code, data, models), used in the paper, properly credited and are the license and terms of use explicitly mentioned and properly respected?
    \item[] Answer: \answerNA{} 
    \item[] Justification: The paper uses no external code, datasets, or models.

\item {\bf New assets}
    \item[] Question: Are new assets introduced in the paper well documented and is the documentation provided alongside the assets?
    \item[] Answer: \answerNA{} 
    \item[] Justification: No new datasets or models are released.

\item {\bf Crowdsourcing and research with human subjects}
    \item[] Question: For crowdsourcing experiments and research with human subjects, does the paper include the full text of instructions given to participants and screenshots, if applicable, as well as details about compensation (if any)? 
    \item[] Answer: \answerNA{} 
    \item[] Justification: No human subjects involved.

\item {\bf Institutional review board (IRB) approvals or equivalent for research with human subjects}
    \item[] Question: Does the paper describe potential risks incurred by study participants, whether such risks were disclosed to the subjects, and whether Institutional Review Board (IRB) approvals (or an equivalent approval/review based on the requirements of your country or institution) were obtained?
    \item[] Answer: \answerNA{} 
    \item[] Justification: No human subjects involved.

\item {\bf Declaration of LLM usage}
    \item[] Question: Does the paper describe the usage of LLMs if it is an important, original, or non-standard component of the core methods in this research? Note that if the LLM is used only for writing, editing, or formatting purposes and does \emph{not} impact the core methodology, scientific rigor, or originality of the research, declaration is not required.
    \item[] Answer: \answerNA{} 
    \item[] Justification: LLM usage was limited to proofreading and formatting and did not affect the core methodology.

\end{enumerate}

\end{document}